\documentclass[showpacs,prl,twocolumn,amsmath,amssymb]{revtex4}
\usepackage{epsfig,epsf,psfrag}
\usepackage{graphicx}
\usepackage{epic,eepic}
\usepackage{color,pstricks}
\begin{document}
\title{Proximity induced interface bound states
in superconductor-graphene junctions}
\date{\today}

\author{P. Burset$^1$, W. Herrera$^2$ and A. Levy Yeyati$^1$}
\affiliation{$^1$Departamento de F\'{\i}sica Te\'orica de la Materia
Condensada C-V, Facultad de Ciencias, Universidad Aut\'onoma de Madrid,
E-28049 Madrid, Spain \\
$^2$Departamento de F\'{\i}sica, Universidad Nacional de Colombia,
Bogot\'a, Colombia}

\begin{abstract}
We show that interface bound states are formed at isolated
graphene-superconductor junctions. These states arise due to the 
interplay of virtual Andreev and normal reflections 
taking place at these interfaces. Simple analytical expressions for
their dispersion are obtained considering
interfaces formed along armchair or zig-zag edges.
It is shown that the states are sensitive to a supercurrent flowing
on the superconducting electrode. The states provide long range 
superconducting correlations on the graphene layer which may be
exploited for the detection of crossed Andreev processes. 
\end{abstract}

\maketitle

{\it Introduction:}
Several striking transport properties have been predicted to
emerge from the peculiar electronic band structure of single atom 
graphite layers, known as graphene \cite{Geim07}. 
Of particular interest is the case of a graphene layer in contact with
a superconducting electrode, a situation which has been explored
in several recent experiments \cite{Heersche07}.
Here, as in the case of normal metals, the
mechanism which dominates the electronic transport at subgap energies
is the Andreev reflection, i.e. the conversion of incident electrons
into reflected holes with the creation of Cooper pairs in the superconductor.
However, while in the case of normal metals the reflected hole has
typically the opposite mean velocity to the incident electron ({\it
retroreflection}) in the case of graphene
it is possible to have Andreev reflections with a {\it specular} 
character, as first shown in \cite{Beenakker06}.

Transport and electronic properties at graphene-superconductor junctions
have been analyzed in several works 
\cite{all-graphene-super}. It has been shown that
the special character of Andreev reflection in graphene leads to 
modifications in the differential conductance compared to that of conventional
N-S junctions \cite{Beenakker06,sengupta-2}. 
The effect on the local density of states (LDOS) has
been studied in Refs. \cite{Tkachov07,burset08}. The properties of bound 
states arising from multiple Andreev reflections in
S-graphene-S junctions have been analyzed in Refs. \cite{titov}.
However, as we show in this work,
the special electronic properties of graphene are such that bound
states can be formed even at {\it isolated} single junctions. 

\begin{figure}
\includegraphics[scale=0.25]{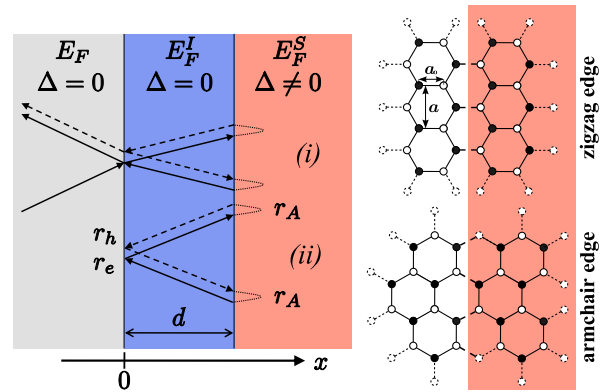}
\caption{(Color online) Simple model for the emergence of IBSs (left panel). It illustrates the
scattering processes taking place at a graphene-superconductor 
interface with an intermediate heavily doped normal graphene region 
of width $d$. 
Cases (i) and (ii) correspond to the case 
$\hbar v q < |E \pm E_F|$ and
$\hbar v q > |E \pm E_F|$ respectively with $\Delta > E > E_F$.
(Right panel) Graphene-superconductor junctions along different edges.
On the superconducting side (shaded areas) the on-site order parameter 
$\Delta$ is finite and the doping level is high ($E^S_{F}\gg \Delta$).}
\label{fig-simple}
\end{figure}

The mechanism for the emergence of these states can
be understood from the scheme depicted in the left panel of 
Fig. \ref{fig-simple}. 
As is usually assumed
the junction can be modeled as an abrupt discontinuity between two
regions described by the Bogoliubov-de Gennes-Dirac equation, taking
a finite superconducting order parameter $\Delta$ and 
large doping $E^S_{F} \gg \Delta$ on the superconducting side and
zero order parameter and small doping $E_F \sim \Delta$ on the
{\it normal} side. For the analysis it is instructive to include
an artificial intermediate normal region with $\Delta=0$ and
$E^I_{F}=E^S_{F}$, whose width, $d$, can be taken to zero at the end of
the calculation. This intermediate region allows to spatially separate
normal reflection due to the Fermi energy mismatch from the Andreev 
reflection associated to the jump in $\Delta$. As shown in 
Fig. \ref{fig-simple} (case i), an incident electron
from the normal side with energy $E$ and parallel momentum $\hbar q$ 
such that $\hbar v q < |E-E_F|$ is partially transmitted into the 
intermediate region and after a sequence of normal and Andreev
reflections would be reflected as a hole.
This process can either correspond to retro or specular 
Andreev reflection depending on whether $E < E_F$ or $E > E_F$
\cite{Beenakker06}.
For $\hbar v q \ge |E\pm E_F|$ neither electron or holes can propagate
within the graphene normal region. However, virtual processes like the
one depicted in Fig. \ref{fig-simple} (case ii) would be present. These correspond 
to sequences of Andreev and normal reflections within the intermediate region. 
A bound state emerges
when the total phase $\phi$ accumulated in such processes reach the resonance
condition $\phi = 2n\pi$. 

The aim of the present work is to demonstrate the existence of 
these interface bound states (IBS) and to analyze their properties
for different types of graphene-superconductor junctions.
After completing the analysis for the case of the simple model
sketched above, which implicitly assumes a decoupling of the two
valleys in the graphene band structure, we
consider more microscopic models for junctions formed along
armchair or zig-zag edges. We study the effect of an additional
potential barrier at the interface and the possibility to 
modify the states by a supercurrent flowing through 
the superconductor. We finally discuss
the potential use of these states for the generation of non-locally 
entangled Andreev pairs. 

It is quite straightforward to determine the 
dispersion relation for the IBS
from the model represented in the left panel of Fig. \ref{fig-simple}. 
The phase accumulated by a sequence
of normal and Andreev reflections in the intermediate region can be obtained
from the corresponding coefficients $r_e$, $r_h$ and $r_A$. 
Following Ref. \cite{Beenakker06} one obtains

\begin{equation}
r_{e,h} = e^{i\alpha^I_{e,h}}
\frac{e^{-i\alpha^{I}_{e,h}} - e^{-i\alpha_{e,h}}}{e^{i\alpha^I_{e,h}}
+ e^{-i\alpha_{e,h}}},
\label{eq-coef-ref}
\end{equation}
where $\alpha^{(I)}_{e,h}=\arcsin{\hbar v q/(E\pm E^{(I)}_F)}$.
The condition $E^I_{F} \gg \Delta, E, 
\hbar v q$ allows to take $\alpha^{I}_{e,h} \simeq 0$. On the other
hand, in the region of evanescent electron and hole states for graphene 
($|\hbar v q|
> |E\pm E_F|$) 
$r_{e,h}$ become a pure phase factor $e^{i\varphi_{e,h}}$, with
$\varphi_{e,h} = -2\mbox{sign}(q/(E\pm E_F))\arctan{e^{\lambda_{e,h}}}$
and $\lambda_{e,h} = \mbox{sign}(q) \mbox{arcosh}(\hbar v q/|E\pm E_F|)$. 
For the Andreev reflection coefficient between regions $I$ and $S$ one
has $r_A = e^{i\varphi_A}$
where $\varphi_A = \arccos{E/\Delta}$, as it corresponds to the Andreev
reflection at an ideal N-S interface with $E^S_F \gg \Delta$ \cite{btk}.
In the limit $d \rightarrow 0$ the total phase accumulated is thus
$\phi = 2 \varphi_A + \varphi_e + \varphi_h$, from which one obtains
the following dispersion relation

\begin{equation}
\frac{E}{\Delta} = \pm 
\frac{e^{(\lambda_e+\lambda_h)/2}-\mbox{sign}(E^2-E_F^2)
e^{-(\lambda_e+\lambda_h)/2}}{2\sqrt{\cosh{\lambda_e}\cosh{\lambda_h}}}.
\label{eq-simple}
\end{equation}

This dispersion simplifies to $E/\Delta = \pm \hbar v q/\sqrt{(\hbar v q)^2 
+ \Delta^2}$ at the charge neutrality point (i.e. for $E_F=0$).
In this case the IBS approaches zero energy
for $q \rightarrow 0$ and tend asymptotically to the superconducting gap
for large $q$. Notice also that the 
decay of the states into the graphene bulk region ($x < 0$ 
in the left panel of Fig. \ref{fig-simple}) is set by $e^{x/\xi_{e,h}}$, where  
$\xi_{e,h} = \hbar v/(|E\pm E_F| \mbox{sinh}(\lambda_{e,h}))$ 
for the electron and hole
components respectively, which can be clearly much larger than
the superconducting coherence length
$\xi_0 = \hbar v/\Delta$ when $E_F \ll \Delta$. It is also interesting
to notice that the IBSs survive when $E_F > \Delta$, i.e. in the regime 
corresponding to the usual Andreev retroreflection,
but with a much smaller spatial extension.

In order to analyze the existence and the characteristics of the IBSs
for different types of graphene-superconductor junctions we make use
of the Green function formalism based on tight-binding models for these
junctions which was introduced in Ref. \cite{burset08}. Within this 
formalism the retarded green functions at the interface 
$\check{\hat{G}}(E,q)$ are given by 
$\left[ \check{\hat{g}}^{-1} - \check{\hat{\Sigma}} \right]^{-1}$, where
$\check{\hat{g}}$ corresponds to the surface of the uncoupled semi-infinite
graphene layer and $\check{\hat{\Sigma}}$ is the self-energy associated
to the coupling with the superconductor. In general all these quantities 
have a $2 \times 2$ structure both in the sublattice (indicated
by the hat symbol) and the Nambu (indicated by the check symbol) spaces.
Once these quantities for each type of interface have
been determined, the existence of an IBS can be established by 
analyzing the equation $ \mbox{det}\left[\check{\hat{g}}^{-1} -
\check{\hat{\Sigma}}\right]= 0$.

{\it Interface along an armchair edge:} 
We first consider an interface constructed along an armchair edge,
as schematically depicted in the right panel of Fig. \ref{fig-simple}.
In a rather generic way one can write $\check{\hat{g}} = 
\hat{g}_e(\check{\tau}_0 + \check{\tau}_z)/2 + 
\hat{g}_h(\check{\tau}_0 - \check{\tau}_z)/2$ 
and $\check{\hat{\Sigma}}/t_g = \beta \check{\tau}_z \check{g}_{BCS} 
\check{\tau}_z + \gamma \check{\tau}_z \hat{\sigma}_x$, where
$\hat{g}_{e,h}$ describe the propagation of $e$ and $h$ components in 
the uncoupled graphene layer, $\check{g}_{BCS} = g \check{\tau}_0 +
f \check{\tau}_x$ with $g = -E f/\Delta = -E/\sqrt{\Delta^2 -E^2}$ being
the BCS dimensionless Green functions, and $\beta$ and $\gamma$ are 
parameters which allow to control the transparency and the type of
interface. As discussed in Ref. \cite{burset08} $\gamma = 0$ corresponds
to a model in which the coherence between the sublattices of 
graphene is broken on the superconducting side (bulk-BCS model), 
whereas for $\beta = \sqrt{3}/2$ and $\gamma=1/2$ corresponds to the
ideal case where superconductivity is induced on the graphene layer
by a superconducting electrode deposited on top,
thus leading to a heavily doped graphene superconductor 
(HDSC). 

To make further analytical progress we take the limit $E,\Delta,\hbar v q
\ll t_g$ in $\hat{g}_{e,h}$ of Ref. \cite{burset08},
where $t_g$ denotes
the hopping element between neighboring sites in the graphene layer. 
In this
case and for $\hbar v q >|E\pm E_F|$, $\hat{g}_{e,h}$ adopt the form 
$t_g \hat{g}_{e,h} = -\frac{1}{2} \left[ \sqrt{3}
\left(\mu_{e,h} \hat{\sigma}_0 + \nu_{e,h} \hat{\sigma}_y\right) 
\pm \hat{\sigma}_x\right]$,
where $\mu_{e,h} = \mbox{sign}(q)/\sinh{\lambda_{e,h}}$ and
$\nu_{e,h} = \mbox{sign}(E\pm E_F)/\tanh{\lambda_{e,h}}$.
The Green functions matrix has the property $\hat{g}^{-1}_{e,h} = 
-t_g^2 \hat{g}_{e,h}^{T} \mp t_g \sigma_x$. Using this property and
the definition for the self-energy the equation for the IBSs in this case
becomes

\begin{eqnarray}
\mbox{det}\left[t_g^2 \hat{g}_h \hat{g}_e + \beta g t_g \left(\hat{g}_e+\hat{g}_h
\right) + \gamma t_g \left(\hat{\sigma}_x \hat{g}_e  - \hat{g}_h \hat{\sigma}_x 
\right) \right. &&\nonumber\\
\left. - (\beta^2 + \gamma^2) \right]= 0. &&
\end{eqnarray} 

\begin{figure}
\includegraphics[scale=0.25,angle=90]{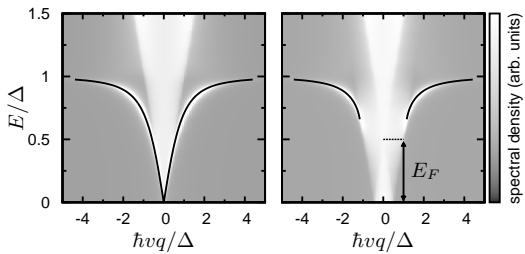}
\caption{Gray-scale plot of the spectral density at a distance
$\sim \xi_0$ from the interface defined along an armchair edge. The results
were obtained using the HDSC model of Ref. \cite{burset08} for 
$E_F=0$ (left panel) and $E_F=\Delta/2$ (right panel). The full lines
indicate the position of the IBS determined from Eq. (\ref{eq-simple}).}
\label{fig-arm}
\end{figure}

For the HDSC model (i.e. $\beta = \sqrt{3}/2$ 
and $\gamma=1/2$) the equation for the IBSs reduce to
the one already found within the simple analytical
model (Eq. (\ref{eq-simple})). This leads to a single root for arbitrary 
doping which is four-fold degenerate due to valley and spin
symmetry.
Fig. \ref{fig-arm} shows a color-scale plot of the spectral density at
a distance $\sim \xi_0$ from the interface on the graphene layer with
two different doping conditions. The full lines correspond to the
IBS dispersion obtained by solving Eq. (\ref{eq-simple}). As can be observed,
the minimal energy for the IBSs, $E_{min}$, depends on $E_F$. 
Further analysis of Eq. (\ref{eq-simple}) reveals that it satisfies the
cubic equation $E_{min}^3 + E_{min}^2 E_F - \Delta^2 E_F=0$, thus
evolving between 0 and $\Delta$ as $E_F$ increases. The transition
between $E_{min}>E_F$ and $E_{min}<E_F$ occurs at $E_F=\Delta/\sqrt{2}$.
The presence of the IBSs manifests also in the appearance of singularities
in the LDOS around $E=\pm\Delta$ (see Ref. \cite{burset08}).
The behavior of the LDOS is analyzed in more detail below.

On the other hand, for the bulk-BCS model (i.e. $\gamma = 0$ and 
$\beta \in (0,1)$) one obtains 
\begin{eqnarray}
\frac{3}{2} \beta^2 g^2 (\mu_e \mu_h + \nu_e \nu_h - 1)  
+ \sqrt{3} \beta g (\mu_e + \mu_h) (1 + \beta^2) & & \nonumber \\
+ \frac{\beta^2(1+2\beta^2)}{2} + \frac{3}{4}(1+2\beta^2)
(\nu_e \nu_h - \mu_e \mu_h) + \frac{1}{4} = 0. &&
\end{eqnarray}

In this case the degeneracy associated to the two valleys in the 
band-structure of graphene is generally broken (except for $E_F=0$).
The roots gradually evolve towards the linear dispersion 
$|E+E_F|=\hbar v q$ as $\beta \rightarrow 0$, which corresponds
to the armchair edge state of the isolated graphene layer 
\cite{sengupta06}.

\begin{figure}
\includegraphics[scale=0.25,angle=90]{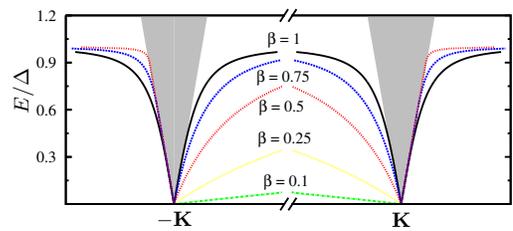}
\caption{(Color online) Dispersion relation for the IBSs on a zig-zag 
interface for decreasing parameter $\beta$ 
controlling the coupling with the superconductor. The parallel momentum
$q$ in Eq. (\ref{eq-zz}) is measured from the Dirac points at $K = \pm2\pi/3a$.}
\label{fig-zz}
\end{figure}

{\it Interface along a zig-zag edge:} 
We now consider an interface along a zig-zag edge as illustrated in 
the right panel of Fig. \ref{fig-simple}.
The Green functions for the semi-infinite zig-zag edge can be obtained
following the same formalism as in Ref. \cite{burset08}. In the continuous
limit $\hat{g}_{e,h}$ becomes

\begin{equation}
t_g \hat{g}_{e,h} = \left( \begin{array}{cc} ie^{-i\alpha_{e,h}} & \mp e^{i\pi/3} \\
\mp e^{-i\pi/3} & 0 \end{array} \right),
\end{equation} 
where as in Eq. (\ref{eq-coef-ref}) 
$\sin \alpha_{e,h} = (\hbar v q)/(E\pm E_F)$ but with
$q$ measured with respect
to the point ${\bf K} = 2\pi/3a$, where $a$ is the lattice constant
indicated in the right panel of Fig. \ref{fig-simple}. There exists an additional
branch where $q$ is measured from the opposite Dirac point at $-{\bf K}$.
The self-energy due to the coupling with the superconductor is in this case
$\check{\hat{\Sigma}} = \beta t_g (\hat{\sigma}_0 + \hat{\sigma}_z) 
\left(\check{\tau}_z \check{g}_{BCS} \check{\tau}_z \right)/2$.
The equation for the IBSs then becomes

\begin{equation}
\frac{E}{\Delta} = \pm \frac{e^{(\lambda_e+\lambda_h)/2} - 
\mbox{sign}(E^2-E^2_F)
\beta^2 e^{-(\lambda_e+\lambda_h)/2}}
{\sqrt{(e^{\lambda_e} + \beta^2 e^{-\lambda_e}) 
(e^{\lambda_h} + \beta^2 e^{-\lambda_h})}}, 
\label{eq-zz}
\end{equation}
which looks very similar to Eq.(\ref{eq-simple}) except for the presence
of the parameter $\beta$ controlling the coupling and the already mentioned
redefinition of the parallel momentum $q$.
An interesting property of zig-zag edges 
is the presence of zero
energy states for total parallel momentum between $(-{\bf K},{\bf K})$ and
$E_F=0$ \cite{saito}. 
When the coupling to the superconductor is turned on by 
increasing the parameter $\beta$, one
observes that the zero energy states evolve acquiring a finite
slope. These states can thus be identified with the IBS for this type of
interface.
This is illustrated in Fig. \ref{fig-zz}. 
When the coupling parameter $\beta$ reaches 1 the usual dispersion of the
simplest analytical model is recovered. 

\begin{figure}
\includegraphics[scale=0.25,angle=90]{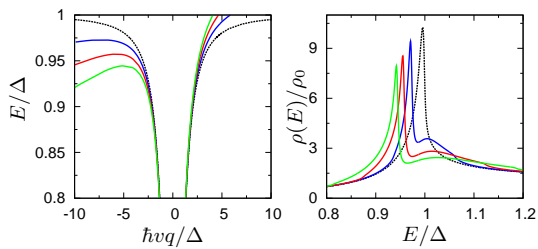}
\caption{(Color online) Effect of a supercurrent flowing on the superconducting
electrode on the dispersion relation (left panel) and on the local
density of states at a distance $\sim \xi_0/10$
from the interface normalized to $\rho_0=\Delta(a/\hbar v)^2/2\pi$
(right panel). 
The results correspond to 
$\hbar v q_s/\Delta = 0.0, 0.25, 0.5$ and $0.75$ with $E^S_F = 100 \Delta$.
}
\label{fig-supercurrent}
\end{figure}

{\it Effect of a supercurrent:} a supercurrent flowing on the superconducting
side of the junction modifies the spatial variation of the phase 
of the order parameter which produces a Doppler shift in the energy of the quasiparticles. This shift, obtained from the Bogoliubov-de Gennes-Dirac equations within the Andreev approximation, is given by $\eta = (\hbar v)^2 q_s q/E^S_{F}$, where $\hbar q_s$ is the momentum of the Cooper pairs assumed to be parallel to the interface. This result is equivalent to the one found in Refs. \cite{zhang04} for conventional and two-band superconductors. Notice that for this analysis we go beyond the limit $E^S_{F} \rightarrow \infty$ taken in the initial simple model. The expression
for the reflection coefficients of Eq. (\ref{eq-coef-ref}) still holds
but $\alpha^I_e \simeq -\alpha^I_h \equiv \alpha^I=\arcsin{\hbar v q/E^S_F}$ 
is kept finite. 
On the other hand, the phase of the Andreev reflection coefficient
between the intermediate region and the current-carrying superconductor
becomes $\varphi_A = \arccos{E^{\prime}/\Delta(q_s)}$, where $E^{\prime} = E+ \eta$. 
At zero temperature, due to Landau criterion, the order parameter is 
unaffected by the supercurrent while $\hbar v q_s \lesssim 
\Delta\left( 0 \right)$ \cite{zhang04}. Therefore, in this condition, 
$\Delta(q_s) \simeq \Delta(0) \equiv \Delta$.
For the case $E > E_F$ we thus get the following modified equation
for the IBSs within the simple model sketched in 
Fig. \ref{fig-simple}

\begin{equation}
\frac{E^{\prime}}{\Delta}= \pm \frac{\sinh{(\lambda_e+\lambda_h)/2}
+ \sin{\alpha^I} \sinh{(\lambda_e-\lambda_h)/2}}{\sqrt{(\cosh{\lambda_e}
- \sin{\alpha^I})(\cosh{\lambda_h} + \sin{\alpha^I})}}.
\end{equation}
Figure \ref{fig-supercurrent} illustrates the effect of a supercurrent
both in the dispersion relation of the IBS (left panel) and in the
local density of states close to the interface (right panel).
For $q_{s}=0$, the IBS manifest in a finite LDOS for $E<\Delta$ and a
sharp peak at $E=\Delta$.
Qualitatively, the presence of a supercurrent breaks the symmetry with
respect to inversion of the parallel momentum $\hbar q$ and leads to a 
splitting of the singularity at $E \simeq \Delta$ in the LDOS. Note that this implies the appearance of an induced net current on the graphene side (for $\left|x \right| \lesssim \xi$).
For $E_F=0$ the distortion of the dispersion relation for 
finite and small $q_s$ is given by $E(q_s,q)=E(0,q)+
(\hbar v q)^2 \eta/((\hbar v q)^2 + \Delta^2)$.

A more quantitative analysis of the effect of a supercurrent
requires the estimation of the parameter
$E^S_{F}$. This parameter is very much dependent on the fabrication
methods and material properties of the metallic electrodes deposited
on top of the graphene layer. According to the ab-initio calculations
of Ref. \cite{cuniberti} for Pd on graphene a typical estimate would
be $E^S_{F} \sim 0.1 eV$, which for a superconductor like Nb
gives a ratio $E^S_{F}/\Delta \sim 100$. The results on 
Fig. \ref{fig-supercurrent} have been obtained for this ratio. 

{\it Conclusions:} 
We have shown that interface bound states appear at
graphene-superconductor junctions. The properties of these states
are sensitive to the type of edge forming the interface, its transparency
and the doping conditions of the graphene layer. 
We have demonstrated that the interface states evolve towards the edge
states of the isolated graphene layer when the transparency 
of the interface is reduced. 
We have also shown that
they can be modulated by a supercurrent flowing through the superconductor
in the direction parallel to the interface.  
Even when our analysis has been restricted to interfaces along
armchair or zig-zag edges we expect the appearance of IBSs to be a general property of any edge
orientation. We also notice that inclusion of weak disorder along the interface introducing a small uncertainty in the parallel momentum $\delta q$ would not prevent the emergence of IBSs provided that $\delta q \ll \Delta/\hbar v$.

As a final remark we would like to comment that 
the existence of IBSs induce long range superconducting 
correlations between distant points on the graphene layer that are
close to the interface. This property could be exploited to detect
crossed Andreev processes and therefore entangled electron pairs 
using weakly coupled STM probes on a graphene-superconductor
junction, in a configuration like the one proposed in Ref. \cite{flatte}. 
The analysis of non-local transport in this system
will be the object of a separate work. 

\acknowledgments
The authors would like to thank correspondence with C.W.J. Beenakker
and useful discussions with J.C. Cuevas and A. Mart\'{\i}n-Rodero.
Financial support from Spanish MICINN under
contracts FIS2005-06255 and FIS2008-04209, by DIB from Universidad
Nacional de Colombia, and by EULA-Nanoforum is acknowledged.

\end{document}